\documentclass{rspublic}
\usepackage{graphicx}

\begin{document}

\title[GRB\,030329 Radio Afterglow Monitoring]{GRB\,030329: Three Years of Radio Afterglow Monitoring}

\author[A.J.~van~der~Horst~and~others]{A.J.~van~der~Horst$^{\,1}$, A.~Kamble$^{\,2}$, R.A.M.J.~Wijers$^{\,1}$, L.~Resmi$^{\,2,3}$, D.~Bhattacharya$^{\,2}$, E.~Rol$^{\,4}$, R.~Strom$^{\,5,1}$, C.~Kouveliotou$^{\,6}$, T.~Oosterloo$^{\,5}$, C.H.~Ishwara-Chandra$^{\,7}$}

\affiliation{$^{1}$Astronomical~Institute, University~of~Amsterdam, Kruislaan~403, 1098~SJ~Amsterdam, The~Netherlands.\\
$^{2}$Raman~Research~Institute, Bangalore~560080, India.\\
$^{3}$Joint~Astronomy~Programme, Indian~Institute~of~Science,\\ Bangalore~560012, India.\\
$^{4}$Department~of~Physics~and~Astronomy, University~of~Leicester,\\ University~Road, Leicester~LE2~7RH, UK.\\
$^{5}$ASTRON, P.O.~Box~2, 7990~AA~Dwingeloo, The~Netherlands.\\
$^{6}$NASA/MSFC, NSSTC, VP62, 320~Sparkman~Drive,\\ Huntsville, AL~35805, USA.\\
$^{7}$National~Center~for~Radio~Astrophysics, Post~Bag~3,\\ Ganeshkind, Pune~411007, India.}

\label{firstpage}

\maketitle

\begin{abstract}{Gamma-ray bursts, radio afterglows, afterglow modeling}
Radio observations of gamma-ray burst (GRB) afterglows are essential for our understanding of the physics of relativistic blast waves, as they enable us to follow the evolution of GRB explosions much longer than the afterglows in any other wave band. We have performed a three-year monitoring campaign of GRB\,030329 with the Westerbork Synthesis Radio Telescopes (WSRT) and the Giant Metrewave Radio Telescope (GMRT). Our observations, combined with observations at other wavelengths, have allowed us to determine the GRB blast wave physical parameters, such as the total burst energy and the ambient medium density, as well as investigate the jet nature of the relativistic outflow. Further, by modeling the late-time radio light curve of GRB\,030329, we predict that the Low-Frequency Array (LOFAR, 30-240~MHz) will be able to observe afterglows of similar GRBs, and constrain the physics of the blast wave during its non-relativistic phase.
\end{abstract}

\section{Gamma-ray burst radio afterglows}
Radio light curves of GRB afterglows display a very different behaviour compared to their optical or X-ray light curves (Frail et al. 1997). The reason for this is the shape and evolution of the broadband synchrotron spectrum. The broadband spectrum is characterised by its peak flux and the three break frequencies, namely the peak frequency $\nu_{\,\mbox{m}}$ that corresponds to the minimal energy in the electron energy distribution, the cooling frequency $\nu_{\,\mbox{c}}$ that corresponds to electrons that cool on the dynamical timescale, and the synchrotron self-absorption frequency $\nu_{\,\mbox{a}}$. These break frequencies and the peak flux can be described in terms of the energy in the blast wave, the density of the surrounding medium, and the fractional energy in the radiating electrons and in the downstream magnetic field (Wijers \& Galama 1999). These four parameters can only be determined if the evolution of all three break frequencies and the peak flux is known, i.e. when well-sampled light curves at X-ray, optical and radio wavelengths are at hand. 

In most afterglows $\nu_{\,\mbox{m}}$ shifts to between the radio and optical regimes within a few hours, which is the usual timescale for the start of radio afterglow observations, by when $\nu_{\,\mbox{a}}$ lies in the radio wave band. Therefore the evolution of $\nu_{\,\mbox{m}}$ and $\nu_{\,\mbox{a}}$ can only be studied at radio wave bands. However, the synchrotron self-absorption causes a steep spectral decline at long wavelengths, making it difficult to detect radio afterglows at early times. At later times, as the peak of the broadband spectrum moves towards the radio regime, the radio light curve shows a rise, in contrast with the declining light curves at optical and X-ray frequencies. Variations in the temporal slope are caused by the passages of $\nu_{\,\mbox{m}}$ and $\nu_{\,\mbox{a}}$ through the observing bands, and by changes in the dynamics of the blast wave. These changes occur when the relativistic blast wave starts spreading laterally, manifesting as a ``jet-break'' in optical and X-ray light curves, and when the blast wave becomes non-relativistic.

\section{Gamma-ray burst 030329 afterglow}
GRB\,030329 displayed one of the brightest afterglows ever, enabling the study of its evolution for a long time and in detail over a broad range of frequencies, from X-ray to centimetre wavelengths. The afterglow was still visible in radio waves 1100 days after the burst trigger. The optical afterglow was visible for only a couple of weeks until it was obscured by the emerging supernova associated with the GRB (Hjorth et al. 2003, Stanek et al. 2003); the X-ray afterglow was detected for only a few months (Tiengo et al. 2004). Thus using our radio observations we can follow the evolution of the blast wave into the non-relativistic phase and uniquely determine the physical parameters of the blast wave.

The broadband afterglow can be modelled by the standard relativistic blast wave model, assuming either a `double jet model' (Berger et al. 2003) or a `refreshed jet model' (Resmi et al. 2005). In both the models, the early-time optical and X-ray light curves are explained by a jet with a small opening angle. The double jet model assumes a co-aligned wider jet component that carries the bulk of the energy and produces the later-time light curves. In the refreshed jet model, the initial jet is `re-energised' by the central engine during its lateral expansion, to make it collimated to a wider opening angle. This refreshed jet then produces the late-time emission. Because the peak and self-absorption frequency of the broadband spectrum of either the wide jet component or the refreshed jet are situated in the centimetre regime, this component is naturally best studied at these wave bands.

\section{Results of the Westerbork Synthesis Radio Telescope and Giant Metrewave Radio Telescope}
\begin{figure}
\begin{center}
\includegraphics[height=0.55\textheight]{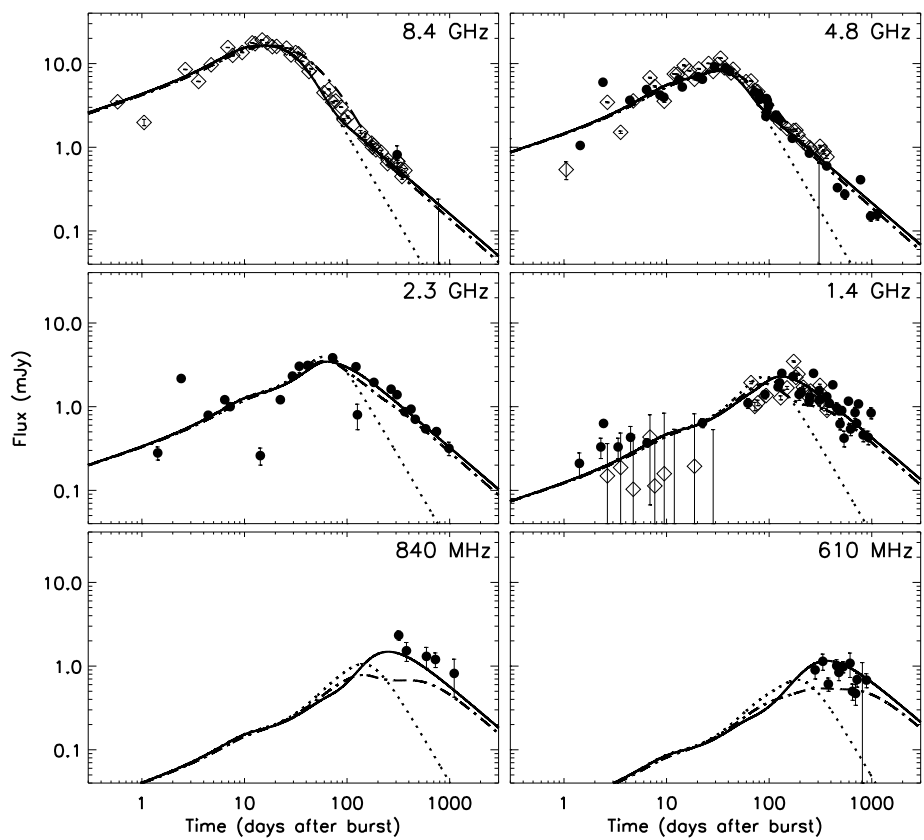}
\end{center}
\caption{Modeling results of the afterglow of GRB\,030329 at centimetre wavelengths. Our light curves obtained with WSRT and GMRT are shown together with earlier reported fluxes from WSRT (Van der Horst et al. 2005), GMRT (Resmi et al. 2005), and VLA \& ATCA (Berger et al. 2003, Frail et al. 2005). The filled circles are WSRT \& GMRT measurements, the open diamonds are VLA \& ATCA measurements. Three fits to the data are shown: the dotted line represents a fit to the first 100 days of radio observations with a wide jet expanding in a homogeneous medium; the solid line corresponds to a model in which the blast wave becomes non-relativistic after 80 days; the dash-dotted line corresponds to a model in which a third jet-component with an even wider opening angle is present. The latter model is excluded by the observations below 1~GHz, which leaves the model with the non-relativistic phase after 80 days as the preferred model for the late-time behaviour of the blast wave.\label{fig:1}}
\end{figure}

The afterglow of GRB\,030329 was observed with the WSRT and the GMRT from 325~MHz to 8.4~GHz, spanning a time range of 268 to 1128 days after the burst (Van der Horst et al. 2006), see figure \ref{fig:1}. We have modelled the light curves we obtained together with earlier reported fluxes from WSRT (Van der Horst et al. 2005), GMRT (Resmi et al. 2005), and VLA \& ATCA (Berger et al. 2003, Frail et al. 2005). The afterglow was clearly detected at all frequencies except for 325~MHz, where we could only obtain upper limits (1-$\sigma$ limits of $\sim 0.5$ mJy at $\sim 2$ years after the burst). GRB\,030329 is the first afterglow to be detected at frequencies below 1~GHz: at 840~MHz with WSRT and even as low as 610~MHz with GMRT. 

The light curves show the peak of the broadband synchrotron spectrum moving to lower frequencies in time. After 80 days the observed light curves decrease less steeply than expected, which can be explained by a transition into the non-relativistic phase of the blast wave (Van der Horst et al. 2005, Resmi et al. 2005, Frail et al. 2005). It was suggested in Van der Horst et al. 2005 that this late-time behaviour could also be explained by a third jet-component with an even wider opening angle than the first two. However, the latter model is excluded by the observations below 1~GHz, which leaves the model with the non-relativistic phase after 80 days as the preferred model for the late-time behaviour of the blast wave. The precise value of the time of the transition into the non-relativistic phase varies between 40 to 80 days for the different modelling methods applied in Van der Horst et al. 2005, Resmi et al. 2005, and Frail et al. 2005.

Our modeling method also allows us to calculate the source size and its evolution (Van der Horst \& Wijers 2006), and from that derive the scintillation properties of this afterglow. Scintillation due to the local interstellar medium modulates the radio flux of the afterglow, as can be seen in figure \ref{fig:1}. Van der Horst et al. (2005) suggested that the early ($\sim$1.5 days) peak that seems to be present in the radio light curves could be the signature of the narrow jet-component causing the early optical and X-ray light curves. They concluded that interstellar scintillation could be a second and better explanation, because of the modeling constraints of the narrow jet-component from the optical and X-ray observations. However, our calculations of the source size and the expected scintillation cannot account for the observed fluxes at early times. Resmi et al. (2005) pointed out that the deceleration of the wide jet-component is expected to be accompanied by a radio flash from the reverse shock, as was also predicted by Piran et al. (2003). This could be the explanation for the observed fluxes.

\section{Low-Frequency Array}
The Low-Frequency Array will be a major new multi-element, interferometric, imaging telescope designed for the 30-240~MHz frequency range. LOFAR will use an array of simple omni-directional antennas. The electronic signals from the antennas are digitised, transported to a central digital processor, and combined in software to emulate a conventional antenna. LOFAR will have unprecedented sensitivity and resolution at metre wavelengths. This will give the GRB community the opportunity to study bright afterglows on even longer timescales than with observations at centimetre wavelengths. 

\begin{figure}
\begin{center}
\includegraphics[height=0.45\textheight]{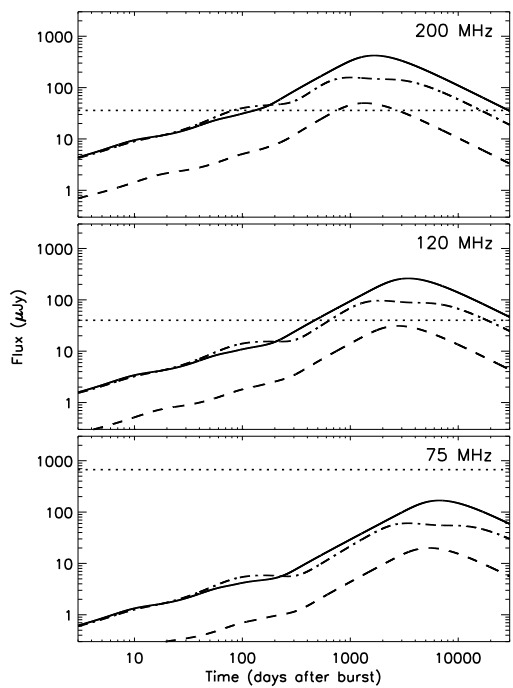}
\end{center}
\caption{The predicted light curves of GRB\,030329 at three frequencies within the LOFAR observing range. The solid line and dash-dotted line correspond to the two models shown in figure \ref{fig:1}; the dashed line shows the predicted light curve of GRB 030329 when situated at a redshift of 1 instead of 0.16. The horizontal dotted lines are the sensitivity limits after 4 hours of observing. It shows that GRB afterglows similar in brightness to that of GRB 030329 can easily be detected with LOFAR on timescales of months to decades; and that fainter afterglows can also be detected after long integration times.\label{fig:2}}
\end{figure}

We have extrapolated the modeling results of the radio afterglow of GRB\,030329 to the LOFAR observing range (Van der Horst et al. 2006), see figure \ref{fig:2}. The predicted light curves show that GRB\,030329 will be observable in the high band of LOFAR (120-240~MHz), but not in the low band (30-80~MHz). The light curves are expected to peak in 2009, when LOFAR will be fully operational, and even later going down from 240 to 120~MHz. We also calculated light curves for GRB\,030329 if it were situated at a redshift of 1 instead of 0.16. The resulting fainter afterglow can also be detected, although with longer integration times, i.e. on the order of a day instead of an hour.

\begin{acknowledgements}
We greatly appreciate the support from the WSRT and GMRT staff in their help with scheduling these observations. The Westerbork Synthesis Radio Telescope is operated by the ASTRON (Netherlands Foundation for Research in Astronomy) with support from the Netherlands Foundation for Scientific Research (NWO). The Giant Metrewave Radio Telescope is operated by the National Center for Radio Astrophysics of the Tata Institute of Fundamental Research.
\end{acknowledgements}

\end{document}